\documentclass[a4paper]{article}

\usepackage{INTERSPEECH2020}

\title{Enhancing Generalization in Audio Deepfake Detection: A Neural Collapse based Sampling and Training Approach}
\name{Mohammed Yousif$^{1}$, Jonat John Mathew$^{1}$,
       Huzaifa Pallan$^{1}$,
       Agamjeet Singh Padda$^{1}$,
       Syed Daniyal Shah$^{1}$,
       Sara Adamski$^{1}$,
       Madhu Reddiboina$^{1}$,
       Arjun Pankajakshan$^{1}$
       }
\address{$^1$ RediMinds Research, USA}
\email{\{firstname\}.\{lastname\}@rediminds.com}

\begin{document}

\maketitle
\begin{abstract}
Generalization in audio deepfake detection presents a significant challenge, with models trained on specific datasets often struggling to detect deepfakes generated under varying conditions and unknown algorithms. While collectively training a model using diverse datasets can enhance its generalization ability, it comes with high computational costs. To address this, we propose a neural collapse-based sampling approach applied to pre-trained models trained on distinct datasets to create a new training database. Using ASVspoof $2019$ dataset as a proof-of-concept, we implement pre-trained models with Resnet and ConvNext architectures. Our approach demonstrates comparable generalization on unseen data while being computationally efficient, requiring less training data. Evaluation is conducted using the In-the-wild dataset.
\end{abstract}


\section{Introduction}
Deepfake audio detection involves classifying audio recordings as either real (authentic/bonafide) or fake (counterfeit/spoof) using deep learning techniques \cite{balamurali2019toward, alzantot2019deep, almutairi2022review}. A key challenge in this field is ensuring that models generalize well to unseen data conditions. This generalization ability can be hindered by \emph{within-class variability} among fake audio samples, which are generated using diverse methods such as text-to-speech (TTS) \cite{kaur2023conventional}, voice cloning \cite{arik2018neural}, and voice conversion \cite{kinnunen2018spoofing} technologies. Additionally, training instability stemming from data imbalance poses another obstacle. Typically, datasets used for training contain a large proportion of fake samples generated using various algorithms compared to real audio samples. This data imbalance issue is prevalent in widely-used datasets like ASVspoof $2019$ \cite{todisco2019asvspoof}, FoR \cite{reimao2019dataset}, and Wavefake \cite{frank2021wavefake}.

Efforts have been made to investigate the generalization of audio deepfake detection models \cite{muller2022does, kawa2022attack, xie2023learning, chen2020generalization}. The common issue lies in the poor detection performance of a deepfake audio detection model trained on one dataset when applied to another dataset, as analyzed in \cite{muller2022does}. An intuitive approach to improve generalization involves training models on diverse datasets, as explored in \cite{kawa2022attack}. However, the data splitting and sampling approach proposed in this study is time-consuming and not autonomous. Other techniques, such as robust feature representation using self-supervised learning \cite{xie2023learning}, and robust feature embedding with improved loss functions and frequency masking \cite{chen2020generalization}, have also been proposed to enhance generalization in deepfake audio detection models.

In this study, we employ the \emph{neural collapse} \cite{papyan2020prevalence, kothapalli2023neural} theory as a data sampling method for training audio deepfake models. Neural collapse characterizes certain properties observed in the penultimate layer embedding of a deep neural classifier during its final stages of supervised training. One such characteristic is \emph{variability collapse}, where within-class variability diminishes as training progresses, leading to data samples of the same class converging around their class mean in the penultimate embedding. Another observed property in neural collapse theory is the capability to assign an unseen data sample to its nearest neighbor class during inference. This assignment is based on the penultimate feature of the sample in relation to the class means computed from the penultimate embeddings of training samples. The neural collapse principles have been previously applied in various domains, including transferability \cite{suresh2023pitfalls}, generalization \cite{galanti2022on}, and interpretability analysis \cite{papyan2020prevalence} of deep classifiers.  

In this study, we leverage the variability collapse property of neural collapse to formulate a sampling approach for selecting representative data points from diverse datasets, thereby creating a new training database. This method exploits the discriminative characteristics observed in the penultimate embedding to identify confidently classified real and fake samples by a pre-trained deepfake audio model. We demonstrate that deepfake audio models trained on this newly created database exhibit promising generalization across unseen data. Additionally, our approach is computationally efficient, requiring less training data compared to existing methods such as the one proposed in \cite{kawa2022attack}. Another related approach involves using active learning to identify prominent fake samples \cite{lindsey2023reducing}. The subsequent sections provide detailed explanations of the methodology (Section \ref{sec:mtd}), dataset and experimental details (Section \ref{sec:data}), results and discussion (Section \ref{sec:result}), and conclusion (Section \ref{sec:con}).

\section{Methodology}
\label{sec:mtd}
The motivation behind this study is rooted in the straightforward idea that integrating diverse datasets pertaining to the same task during training can improve the generalization capability of a deep learning model. In this study, we apply this concept to the domain of audio deepfake detection. However, a significant challenge in this field arises from the high within-class variability in the fake audio class, primarily due to the use of various algorithms to generate fake audio samples. For instance, in the ASVspoof $2019$ LA dataset \cite{todisco2019asvspoof}, four TTS and two voice conversion algorithms are employed to generate fake audio samples in the training data, resulting in a dataset comprising $22,800$ fake samples and only $2,580$ real samples. Similar diversity in algorithms is observed in other audio deepfake detection datasets such as FoR \cite{reimao2019dataset}, Wavefake \cite{frank2021wavefake}, and In-the-wild \cite{zi2020wilddeepfake}. Consequently, audio deepfake models trained on specific datasets exhibit poor generalization across other datasets, as analyzed in \cite{zi2020wilddeepfake}.

One straightforward approach to addressing the generalization issue is to collectively use diverse audio deepfake datasets and train a large model. While this method can enhance generalization across unseen data distributions, it may increase the variability collapse within the fake class, given the diverse nature of different deepfake algorithms and the varying numbers of fake samples associated with each of them. Additionally, managing such a large training database is computationally intensive and requires meticulous data preprocessing and splitting steps, as demonstrated in \cite{kawa2022attack}. To mitigate these challenges, we propose a neural collapse-inspired data sampling approach in this study. This method involves sampling confidently classified real and fake samples from a deepfake dataset based on the geometric representations of the penultimate embedding of a pre-trained deepfake classifier. Representative real and fake samples are selected using a distance-based scoring function relative to class means.

The methodology employed in this study can be summarized as follows. Initially, we train deepfake audio models on individual datasets. Subsequently, we utilize the pre-trained model to identify correctly predicted real and fake audio samples within the entire training database, termed as \emph{audio samples of interest}, which are then considered for the sampling process. For each pre-trained model, we apply neural collapse principles. This involves extracting the penultimate embedding feature for all audio samples of interest and computing class means associated with each class using Equation \ref{eq:nc}, 
\begin{equation}
\begin{split}
  \mu_k &= \frac{1}{n}\sum_{i=1}^nf_{k,i}
  \label{eq:nc}
\end{split}
\end{equation}
where $n$ represents the total number of samples in class $k$, $f_{k,i}$ denotes the penultimate feature embedding, and $\mu_k$ is the mean of penultimate embeddings for the $k^{th}$ class. Following this, we visualize the geometric distribution of the penultimate embedding of the audio samples of interest alongside the class mean using t-SNE \cite{van2008visualizing}. For each sample in the audio samples of interest belonging to a class, we compute the distance between the sample and the class mean. The sampling approach for each class is based on selecting nearest neighbor samples relative to the class mean. Further details of this sampling approach are outlined in Algorithm~\ref{alg:sam-real}.

This paper presents a proof-of-concept experiment of the proposed methodology using the ASVspoof $2019$ LA dataset. Section \ref{sec:result} outlines a generic version of the proposed method for future research.

\begin{algorithm}[htb!]
\caption{Neural collapse based sampling approach}
\label{alg:sam-real}
\begin{algorithmic}[1]
\renewcommand{\algorithmicrequire}{\textbf{Inputs:}}
\renewcommand{\algorithmicensure}{\textbf{Outputs:}} 
\REQUIRE $\bigl\{\textit{x}_{k,i}\bigr\}_{i=1}^n$ is the set of audio samples of interest belonging to class $k$, $f_{k,i}$ is the penultimate embedding of $\textit{x}_{k,i}$.
\ENSURE sampled set of real audio samples $\bigl\{\textit{x}_{k,i}\bigr\}_{i=1}^m$; $m<n$  
\STATE $\mu_k = \frac{1}{n}\sum_{i=1}^{n} f_{k,i}$
\STATE $d_{k,i} = ||f_{k,i} - \mu_k||$
\STATE \text{sort}$([d_{k,i}])$
\STATE $\bigl\{\textit{x}_{k,i}\bigr\}_{i=1}^m  = \{x_{k,i} \text{ where } d_{k,i} \leq \text{threshold} \}$
\end{algorithmic}
\end{algorithm}

\section{Datasets and Experimental Details}
\label{sec:data}
\subsection{Datasets}
We utilize the ASVspoof $2019$ logical access (LA) \cite{todisco2019asvspoof}, FoR \cite{reimao2019dataset}, and Wavefake \cite{frank2021wavefake} datasets for training audio deepfake models, reserving the In-the-wild \cite{zi2020wilddeepfake} dataset exclusively for evaluation purposes. The ASVspoof LA training dataset contains $2,580$ real samples, based on the VCTK corpus \cite{veaux2017cstr}, and $22,800$ fake samples generated from four TTS and two voice conversion algorithms. The ASVspoof LA evaluation data consists of $7,355$ real and $63,882$ fake samples, generated by seven TTS and six voice conversion spoofing algorithms. The FoR dataset includes $111,000$ real and $87,000$ fake audio samples, sourced from the Arctic \cite{kominek2004cmu}, LJSpeech \cite{ljspeech17}, and VoxForge datasets for real samples, and generated by seven TTS algorithms for fake samples. In the Wavefake dataset, we focus solely on real and fake data points from the LJSpeech dataset, totaling approximately $91,700$ fake samples produced by seven TTS algorithms. The In-the-wild dataset comprises $37.9$ hours of audio, with $17.2$ hours of fake and $20.7$ hours of real data, sourced from publicly available video and audio files. The fake clips in this dataset are created by segmenting $219$ publicly available video and audio files with adversarial perturbations.

\subsection{Experimental Details}
\subsubsection{Pre-trained Models}
We initially utilize the ASVspoof $2019$ LA dataset \cite{todisco2019asvspoof} to develop audio deepfake models, employing ResNet (with $18$ residual blocks) \cite{alzantot2019deep} and ConvNext \cite{liu2022convnet} architectures. These models undergo evaluation using the ASVspoof $2019$ LA evaluation subset and the In-the-wild dataset to highlight the generalization challenges in audio deepfake detection. This experiment is termed as \emph{Experiment $1$}. Subsequently, we train a tiny ResNet-based model (with $9$ residual blocks) using a subset of real and fake samples from the ASVspoof $2019$ LA training database, generated using the sampling method outlined in Section \ref{sec:mtd}. This model is then assessed using the ASVspoof $2019$ LA evaluation dataset and the In-the-wild dataset. By adjusting the threshold value of the distance-based score function, we control the quantity of samples in each class, conducting experiments with varying sampling rates for both the real and fake classes. The entire experiment is termed as \emph{Experiment $2$}. These experiments provide insights into the impact of sampling quantity on model training and generalization.

Additionally, we conduct a parallel experiment by randomly sampling a fixed number of real and fake samples from each of the ASVspoof $2019$ LA, FoR \cite{reimao2019dataset}, and Wavefake \cite{frank2021wavefake} datasets to create a new training database. This process is carried out without utilizing pre-trained models or the sampling method proposed in Section \ref{sec:mtd}. To ensure the absence of duplicate data, especially considering that the LJSpeech corpus serves as the source of real samples in both the FoR and Wavefake datasets, we take care in database creation. We explore various sampling rates, including $1000$, $3000$, and $5000$ samples from both the real and fake classes for each dataset. For each sampling rate and its corresponding newly created training database, we train an audio deepfake model using the tiny ResNet architecture with $9$ residual blocks and evaluate its performance on a subset of the collective database (\emph{eval data} in Table \ref{table:rand-sam}) and the In-the-wild dataset. This experiment termed as \emph{Experiment $3$}, conducted without the need for pre-trained audio deepfake models and with random sampling, aims to analyze the scope of the proposed methodology. Evaluation using the In-the-wild dataset provides insights into the generalization capabilities of each model across unseen data distributions.

\subsubsection{Training and Evaluation}
During model training, we employ audio samples with a duration of $3$ seconds. To ensure uniform input lengths, we apply trimming and padding techniques to adjust long and short audio samples in the datasets, resulting in $3$-second audio samples. Mel-spectrogram features are utilized for training all our models, employing a feature representation of $80$ log mel-bands spectrograms. These features are extracted using a short-term Fourier transform (STFT) with a fast Fourier transform (FFT) window of $512$, a hop length of $160$, and a sample rate of $16$ kHz.

In the Resnet and ConvNext-based models, the activation of convolutional layers undergoes batch normalization and is subjected to regularization with a dropout probability of $0.3$. The weights of the convolutional layers are initialized using a Glorot uniform distribution \cite{glorot2010understanding}. Each model undergoes training for $100$ epochs, utilizing a binary cross-entropy loss function, and the Adam optimizer with a learning rate set at $0.001$. To prevent overfitting, early stopping is employed, terminating training after $10$ epochs based on the validation loss score.

Our primary evaluation metric is the equal error rate (EER), calculated based on the area under the receiver operating characteristic curve (ROC-AUC), which we denote as EER-ROC. This computation method slightly differs from the EER computation based on the detection error tradeoff (DET) curve, as utilized in the ASVspoof challenge \cite{yamagishi2021asvspoof}. Additionally, we utilize mean average precision (mAP) as our secondary evaluation metric. Lower scores in the EER-based metric indicate better model performance, while higher mAP scores signify improved performance. The values for mAP fall within the range of $0$ to $1$.

\section{Results and Discussion}
\label{sec:result}
The results of \emph{Experiment $1$} are presented in Table \ref{table:asv-itw}. Both the Resnet-based and ConvNext-based models demonstrate good performance on the ASVspoof $2019$ evaluation data in terms of EER-ROC and mAP. However, the performance of these models on the In-the-wild dataset is poor. For instance, with the Resnet-based model, the EER-ROC is $0.57$ and mAP is $0.32$. This outcome highlights the generalization issue in audio deepfake models; when trained on a specific dataset, they often fail to generalize across unseen datasets that were not part of the training process.

\begin{table}[t]
\centering
\renewcommand{\arraystretch}{1.5}
\caption{Evaluation performance of the Resnet and ConvNext model using ASVspoof $2019$ evaluation dataset and In-the-wild dataset.}
\label{table:asv-itw}
\begin{tabular}{|cc|c|c|}
\hline
\multicolumn{2}{|c|}{}                                       & Resnet & ConvNext \\ \hline
\multicolumn{1}{|c|}{\multirow{2}{*}{ASV eval}}    & EER-ROC & 0.08   & 0.11     \\ \cline{2-4} 
\multicolumn{1}{|c|}{}                             & mAP     & 0.99   & 0.99     \\ \hline
\multicolumn{1}{|c|}{\multirow{2}{*}{In the wild}} & EER-ROC & 0.57   & 0.47     \\ \cline{2-4} 
\multicolumn{1}{|c|}{}                             & mAP     & 0.32   & 0.41     \\ \hline
\end{tabular}
\end{table}

\begin{figure}[htb!]
  \centering  \includegraphics[width=\linewidth,height=5cm]{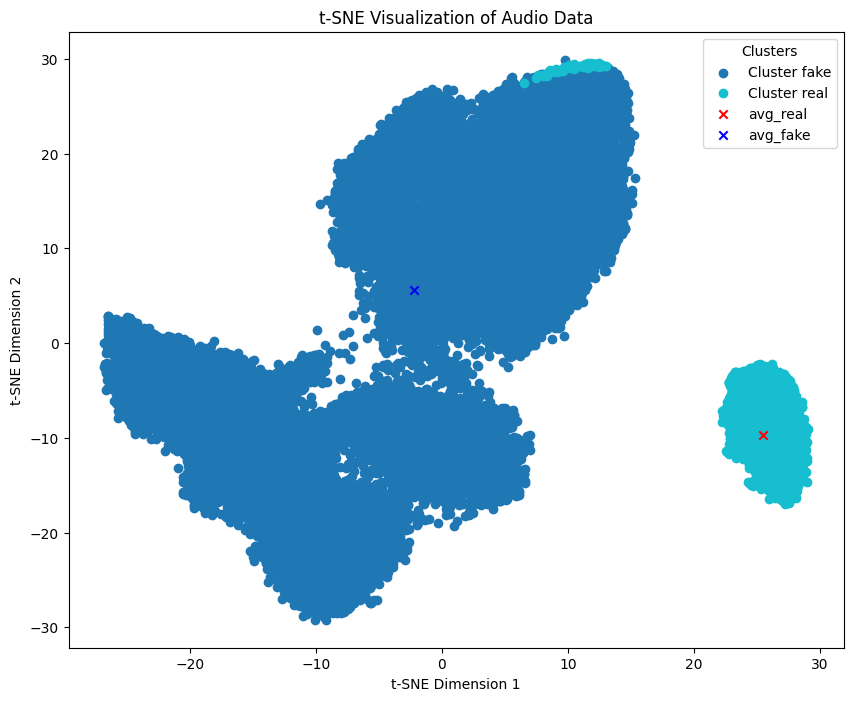}
  \caption{Visualization of the penultimate embedding for real and fake classes in the ASVspoof $2019$ training database.}
  \label{fig:clustering}
\end{figure}

In \emph{Experiment $2$}, we initially examined the penultimate feature embeddings corresponding to both real and fake audio samples across the ASVspoof $2019$ LA training database using the pre-trained Resnet model, as depicted in Figure \ref{fig:clustering}. We empirically applied different sampling rates separately for the real and fake classes based on Algorithm \ref{alg:sam-real}. The optimal result was achieved by utilizing the entire real data and only $50\%$ of the fake data from the ASVspoof $2019$ LA training database. The tiny Resnet model, trained on this new training subset, achieved an EER-ROC of $0.54$ on the In-the-wild dataset and an EER-ROC of $0.10$ on the ASVspoof $2019$ LA evaluation dataset. Additionally, in terms of mAP, the model achieved $0.48$ on the In-the-wild dataset and $0.99$ on the ASVspoof $2019$ LA evaluation dataset. The outcomes obtained from the ASVspoof $2019$ LA dataset show the potential of our proposed approach. By sampling from large, unbalanced datasets to construct a new training database and subsequently re-training the model, we can improve its generalization capability while minimizing the amount of training data needed.

The outcomes of \emph{Experiment $3$} are depicted in Table \ref{table:rand-sam}. Notably, the model trained with $1000$ samples per class per dataset exhibits the most favorable overall performance. Specifically, this model achieves an EER-ROC value of $0.49$ on the In-the-wild dataset. These results further highlight the potential of our proposed approach.

\begin{table}[t]
\centering
\renewcommand{\arraystretch}{1.5}
\caption{Evaluation performance of the tiny Resnet model using combined evaluation dataset and In-the-wild dataset.}
\label{table:rand-sam}
\begin{tabular}{|cc|c|c|c|}
\hline
\multicolumn{2}{|c|}{\begin{tabular}[c]{@{}c@{}}No. of samples per \\ class per dataset\end{tabular}} & 1000 & 3000 & 5000 \\ \hline
\multicolumn{1}{|c|}{\multirow{2}{*}{eval data}}                        & EER-ROC                     & 0.39 & 0.44 & 0.44 \\ \cline{2-5} 
\multicolumn{1}{|c|}{}                                                  & mAP                         & 0.60 & 0.56 & 0.56 \\ \hline
\multicolumn{1}{|c|}{\multirow{2}{*}{In the wild}}                      & EER-ROC                     & 0.49 & 0.54 & 0.52 \\ \cline{2-5} 
\multicolumn{1}{|c|}{}                                                  & mAP                         & 0.37 & 0.26 & 0.32 \\ \hline
\end{tabular}
\end{table}

We propose a modified version of the sampling algorithm (Algorithm \ref{alg:sam-real}) to specifically sample fake data points as part of our future endeavors. Given the within-class variability issue in the fake class, a variability collapse-based sampling approach, similar to that defined in Algorithm \ref{alg:sam-real}, may not be an appropriate sampling method. Hence, we suggest an alternative method involving k-means clustering \cite{macqueen1967some, sinaga2020unsupervised} on the set of fake samples within the audio samples of interest. The number of clusters can be determined based on the number of algorithms used to generate fake audio samples in the dataset, with the maximum number of clusters set equal to the number of different algorithms associated with the fake class. Various scenarios may arise, such as overlapping clusters due to similarities between algorithms within the fake class. To address this, it is vital to identify the optimal clustering condition by adjusting the number of clusters, aiming to minimize cluster overlap. Subsequently, nearest neighbor sampling can be applied to the samples within each cluster with respect to its cluster center. This approach enables the sampling of a unique set of fake samples associated with each cluster of the fake class, thereby facilitating the elimination of confusing data samples. Further details of this modified sampling approach for the fake class are outlined in Algorithm~\ref{alg:sam-fake}.

In instances where the optimal condition is not observed, two different scenarios are anticipated. Firstly, if we encounter highly overlapping clusters (in the optimal condition) even after the iterative k-means approach, we can select the set of sample data points that adhere to the sampling rule individually and collectively across the two clusters. The sampling approach, as described in Algorithm \ref{alg:sam-real}, is then applied to the remaining non-overlapping clusters. The same procedure can be followed when dealing with situations where more than two clusters exhibit significant overlap. In the worst-case scenario, wherein all clusters overlap during the iterative k-means steps, regardless of the number of clusters, we can sample fake data points that meet the sampling rule individually and collectively across all clusters at the best optimal condition.

\begin{algorithm}[htb!]
\caption{Sampling approach for the fake class}
\label{alg:sam-fake}
\begin{algorithmic}[1]
\renewcommand{\algorithmicrequire}{\textbf{Inputs:}}
\renewcommand{\algorithmicensure}{\textbf{Outputs:}} 
\REQUIRE $\bigl\{\textit{x}_{f,i}\bigr\}_{i=1}^p$ is the set of fake audio samples of interest, $f_{f,i}$ is the penultimate embedding of $\textit{x}_{f,i}$.
\ENSURE sampled set of fake audio samples $\bigl\{\textit{x}_{f,i}\bigr\}_{i=1}^q$; $q<p$ 
\STATE $\textit{S\textsubscript{f}} <- \text{new set to store sampled fake data points}$
\STATE \text{apply k-means clustering on} $\bigl\{f_{f,i}\bigr\}_{i=1}^p$
\FOR{ \text{each cluster} $\textit{c}$}
    \STATE $\mu_c = \frac{1}{r}\sum_{i=1}^{r} f_{f,i}$; $c*r=p$
    \STATE $d_{c,i} = ||f_{f,i} - \mu_c||$
    \STATE \text{sort}$([d_{c,i}])$
    \STATE $\textit{S\textsubscript{f}}^c  = \{x_{f,i} \text{ where } d_{c,i} \leq \text{threshold} \}$
    \STATE $\textit{S\textsubscript{f}}.append(\textit{S\textsubscript{f}}^c)$
\ENDFOR
\RETURN $\textit{S\textsubscript{f}}$
\end{algorithmic}
\end{algorithm}

\begin{figure}[htb!]
  \centering  \includegraphics[width=\linewidth,height=5cm]{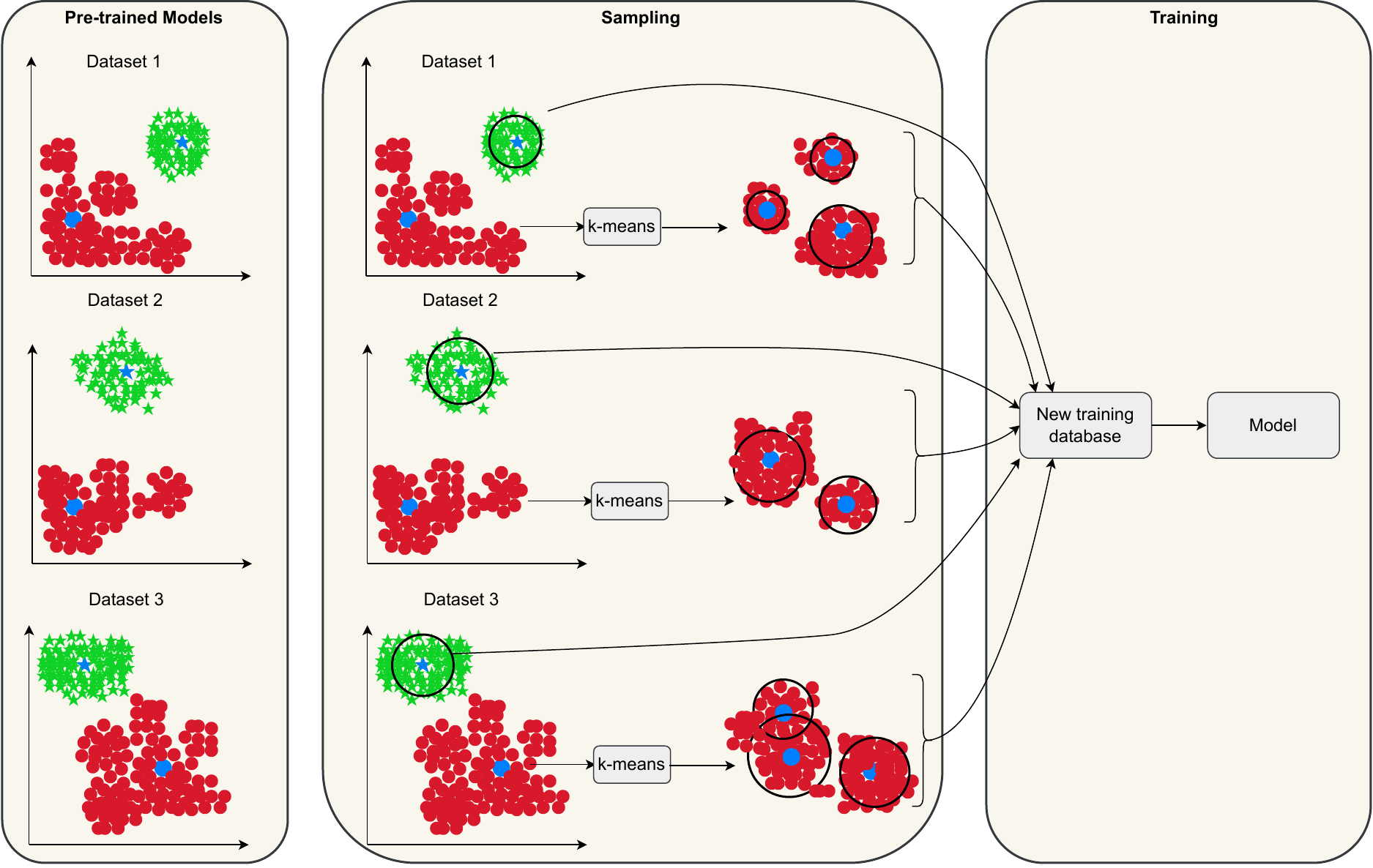}
  \caption{A schematic representation of our proposed methodology. Red data points represent the fake class, green data points represent the real class, and blue data points represent the class means.}
  \label{fig:pipeline_bk}
\end{figure}

The validation of this modified sampling approach is a future work. Utilizing this method, it is possible to sample representative data points from multiple datasets for both the real and fake classes separately. Subsequently, we can merge these sampled data points to construct a new training database for training an audio deepfake model. A schematic representation of the complete process involved in our proposed methodology is depicted in Figure \ref{fig:pipeline_bk}. 

\section{Conclusion}
\label{sec:con}
In conclusion, this work addresses the critical challenge of generalization in audio deepfake detection models across datasets. Initially, we demonstrate this issue by training audio deepfake models using the ASVspoof $2019$ LA dataset and evaluating them on the In-the-wild dataset. Leveraging a neural collapse-based sampling methodology applied to pre-trained models trained on diverse datasets, we propose a methodology to enhance model generalization while reducing the computational burden associated with training on large datasets. Through a series of experiments, we validate the effectiveness of our approach using the ASVspoof $2019$ LA dataset, showcasing comparable performance on unseen data distributions, particularly evident in our tiny Resnet-based model. Our exploration of random sampling techniques also shows the potential of our proposed methodology. Future work involves refining our sampling algorithms, especially tailored for fake data points, to further optimize model training and generalization.

\bibliographystyle{IEEEtran}

\bibliography{mybib}

\end{document}